# Feasibility of Free-Space Transmission using L-Band Maser Signals in Organic Gain Media

Sophia Long[1], Priyanka Choubey[1], Max Attwood[2], Justin Chan[2], Hamdi Torun[1], Juna Sathian[1]

[1]*School of Engineering, Physics and Mathematics, Faculty of Science and Environment, Northumbria University, NE1 8ST, United Kingdom.*

[2]*Department of Materials, Imperial College London, Exhibition Road, London, SW7 2AZ, United Kingdom.*

## Abstract

Atmospheric conditions such as fog, humidity, and scattering by foliage routinely degrade optical free-space (FS) links, motivating alternatives that are robust in adverse conditions. Coherent microwave sources offer a compelling alternative for quantum-secure communication, yet their propagation outside enclosed resonators has remained untested. Here, we demonstrate room-temperature FS transmission of maser signals generated using organic L-band (1-2 GHz) gain media, pentacene doped p-terphenyl (Pc: PTP), and Diazapentacene-doped p-terphenyl (DAP:PTP). Spectral and temporal coherence is preserved over distances up to 25 cm, approximately one wavelength at the masing frequency. Tests included polarisation misalignment, high-humidity conditions, and partial occlusion by foliage to emulate realistic FS reception scenarios. Strong spin-photon coupling was maintained, as confirmed by persistent rabi oscillations and normal-mode splitting. An instantaneous peak output of 4.29 mW (+6.32 dBm) from a 0.01% DAP:PTP gain medium. with a pulse duration of ~4 μs, marks a performance benchmark for directly coupled masers. These findings demonstrate a proof-of-concept for masers as viable platforms for short-range, interference-resilient coherent microwave links, with future relevance to quantum sensing and secure communication technologies.

## Introduction

Wireless communication connects satellites, cellular networks, radar systems, and remote sensing platforms, serving as the backbone of modern information exchange. These systems predominantly operate within the radio frequency (RF) spectrum of the electromagnetic (EM) spectrum (~3 kHz-300 GHz), which are valued for their ability to propagate through clouds, fog, and vegetation, more effectively than optical or infrared signals [1] [2]. However, attenuation can still be significant in dense or wet foliage, where L-band transmission may incur several dB loss per metre depending on water content, density, and path length. As demand for data continues to accelerate, these frequency bands have become densely populated, leading to spectral congestion, interference, and increased risk of crosstalk between overlapping channels [3].

Conventional microwave communication technologies typically employ broadband, incoherent sources that emit noise-prone, omnidirectional radiation [4] [5]. This lack of coherence reduces spectral efficiency, and limits signal fidelity, and limits the density of independent information channels. Emerging quantum technologies require communication systems capable of low-noise, high-fidelity transmission in complex environments. This challenge has created a growing need for signal architectures that are both interference-resilient and compatible with quantum information protocols. Sources with high spectral purity and directional coherence could mitigate cross-channel interference and improve performance in frequency-congested or environmentally challenging conditions. [6] [7] [8].

The MASER, Microwave Amplification by Stimulated Emission of Radiation, offers a fundamentally different approach to microwave signal generation. Unlike classical RF oscillators, masers generate narrowband signals via stimulated emission from an inverted spin ensemble, confined within a resonant cavity [9]. Rather than arising from synthetic modulation, the maser's temporal behaviour (including Rabi oscillations) originates directly from coherent population exchange between spin sublevels [10], governed by the spin–photon coupling strength [11] [12]. Consequently, phase noise is not dominated by circuit modulation but is specified primarily by quantum processes in the gain medium (e.g., spontaneous emission and thermal excitation), with the collective spin-photon coupling scaling as $\sqrt{N}$ for an ensemble of $N$ emitters, as described by the Tavis–Cummings model [13], allowing ensembles to enter the strong-coupling regime and exhibit vacuum Rabi splitting.

The first maser was demonstrated in 1954 using ammonia ($NH_3$) molecules as the gain medium by Gordon, Zeiger, and Townes [14], marking the birth of quantum microwave amplification. The ammonia maser paved the way for the hydrogen maser, developed in 1960 by Goldenberg, Kleppner, and Ramsey, which operates at the 1.42 GHz hyperfine transition of atomic hydrogen and is still used today as ultra-stable frequency references in deep-space tracking and GPS ground stations [15]**.** However, the significant mass (typically over 100 kg [16] [17]), high power consumption, and complex vacuum and magnetic shielding required for hydrogen maser operation have precluded their widespread use onboard satellites, where size, weight, and power (SWaP) constraints are central.

The room-temperature masers were first established in 2012 by optically pumping pentacene-doped p-terphenyl crystals, eliminating cryogenic requirements through tailored organic spin dynamics and resonator design [9]. Building on this foundation, further advances in gain media formulation and mode coupling have now enabled mW peak maser output powers at room temperature (as demonstrated in 2025 [18]), suggesting the practicality and performance of maser-based communication systems. However, all such demonstrations have relied on coaxially coupled or enclosed resonators. FS transmission, utilising the maser's coherence as a powerful quantum source beyond the cavity, remains unexplored.

This work demonstrates FS transmission of a room-temperature coherent maser signal. Using linearly polarised L-band microstrip antennas, maser signal propagation was measured over variable distances from 5 cm to 25 cm, the latter corresponding to approximately one full wavelength ($\lambda \cong 21cm$) at the masing frequencies used (1.45-1.478 GHz). To evaluate link robustness, we performed two controlled perturbations tests. First, receiver antennas were incrementally rotated to assess the system's sensitivity to polarisation misalignment. Second, we introduced controlled environmental conditions, including high humidity and partial occlusion by fresh vegetation, to simulate signal degradation mechanisms that typically impair FS optical communication. FS signal resilience underscores masers as viable sources of spatially and temporally coherent microwave signals in environments where line-of-sight optical systems are impractical. Extension to longer ranges will require higher-gain antennas and a full link budget, but these initial results highlight the feasibility of interference-resilient maser transmission beyond enclosed cavities.

# Results

## 2.1 Maser Setup and Antenna Calibration

To realise room-temperature maser transmission in FS, we first established an experimental platform that integrates an optically pumped organic maser to linearly polarised, L-band microstrip patch antennas with matched polarisation. The platform is designed to support efficient maser generation and FS reception while preserving temporal coherence and minimising coupling and propagation losses.

Characterisation of the loaded maser cavity, gain media, and antenna system is essential to validate baseline performance and guide subsequent transmission experiments. We investigated the microwave resonance behaviour of two pentacene-based organic gain media, Pc: PTP and DAP: PTP, under pulsed optical excitation, as well as the frequency response and angular emission profile of the patch antennas responsible for transmitting and receiving the maser signal. As detailed in the following sections, a high-Q cavity and frequency-matched polarisation aligned antennas provide the foundation for controlled studies of FS propagation, angular sensitivity, and environmental robustness.

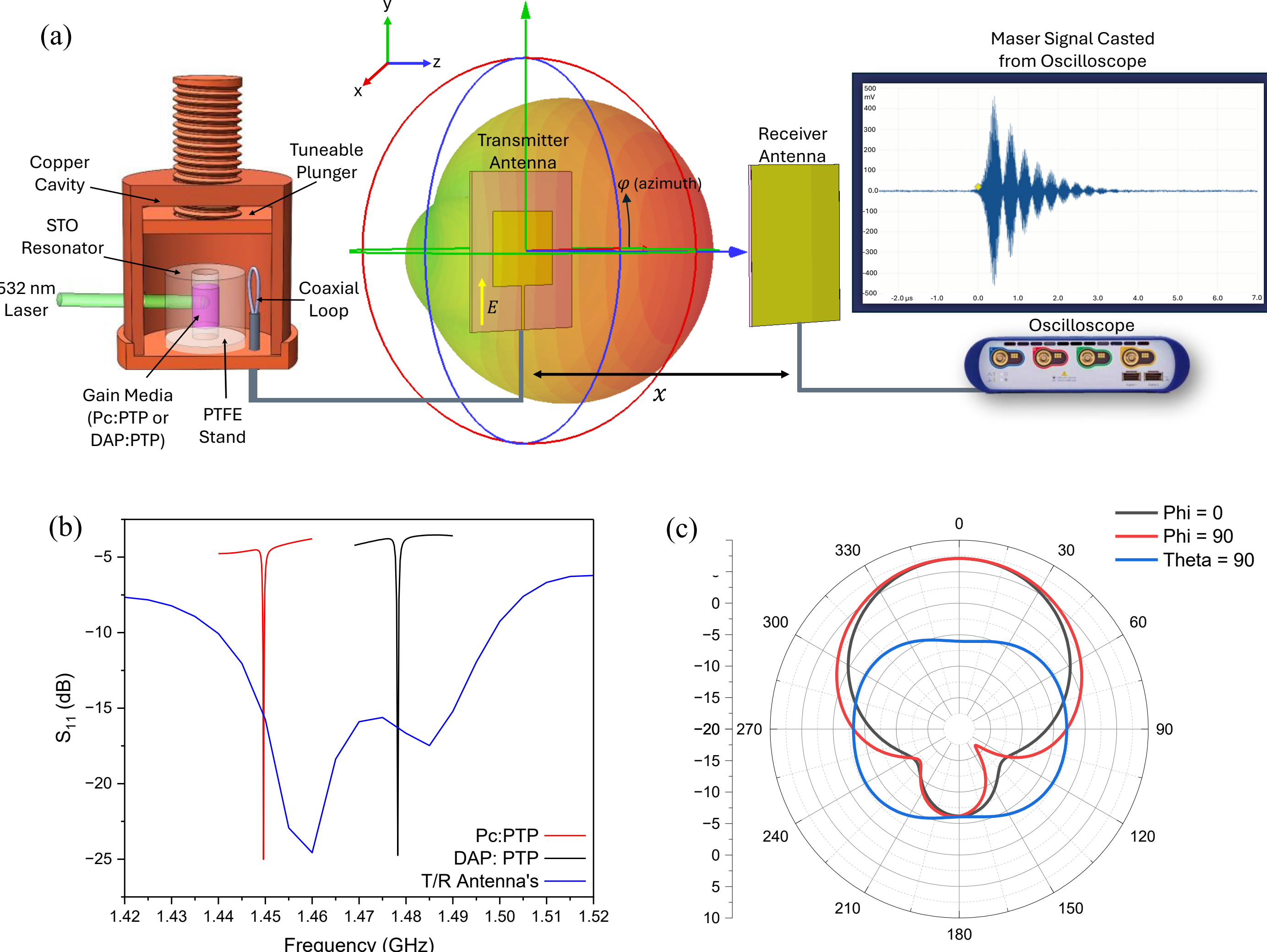

Figure 1: Maser FS setup and antenna analysis. (a) SolidWorks rendering of the experimental setup of the FS maser transmission, illustrating the generation and wireless detection of the maser signal. (b) $S_{11}$ responses of the transmitting/receiving patch antenna (blue), alongside the loaded cavity resonances for Pc:PTP (red) and DAP:PTP (black). The fabricated patch

antenna exhibited minima of –15.8 dB at 1.449 GHz and –16.2 dB at 1.478 GHz, closely overlapping the masing resonances of Pc:PTP and DAP:PTP, respectively. (c) The far-field radiation pattern of the designed patch antenna is simulated in CST Studio Suite. The plot displays gain (in dB) as a function of angle for $\varphi = 0°$ (black), $\varphi = 90°$ (red), and $\theta = 90°$ (blue) where ϕ and θ are azimuthal and elevation angles in spherical coordinates, respectively.

To sustain a FS maser transmission, a high-Q copper cavity incorporating a $SrTiO_3$ dielectric resonator and organic gain media (Pc: PTP at 1.45 GHz or DAP: PTP at 1.478 GHz) was optically pumped with a pulsed 532 nm Nd: YAG laser (Nano S-120-20, Litron Lasers,). The cavity resonance was tuned in frequency by translating a mechanical plunger vertically. Due to the narrow tuning range of masing transitions ($\Delta f \approx$ 3.15 MHz for Pc:PTP and 4.28 MHz for DAP:PTP [18]), custom linearly polarised patch antennas were designed and fabricated in-house, to precisely match the narrow masing frequencies. Antenna parameters were optimised using CST Studio Suite® simulations, while VNA measurements confirmed a measured return loss of $\sim -16 \pm 0.2$ dB at the masing resonance, validating practical impedance matching for both gain media. This return loss is acceptable for proof-of-concept demonstrations, though further improvements (e.g., $S_{11} \approx -20$ to $-25$ dB) are attainable using commercial high-quality antennas.

The antennas exhibit strong linear polarisation and a pronounced main lobe, as confirmed by the far-field simulations (Fig. 1c), providing directional transmission with minimal side lobes to maximise signal strength.

## 2.2 Free-Space Maser Signal Performance

To establish a quantitative baseline for all subsequent tests, we first verified the transmission performance of the room-temperature maser platform under controlled FS conditions. This experiment assessed whether the maser output was sufficiently strong to overcome the initial dielectric and propagation losses, and to confirm that the transmitter-receiver pair functioned reliably for both Pc:PTP and DAP:PTP samples. Measurements of maser peak output, signal-to-noise ratio (SNR), and Rabi frequency were recorded as a function of distance between direct connection (0 cm) to 25 cm. A detectable maser signal was observed across the full range of both gain media, establishing the baseline link performance for later environmental analysis. In addition, frequency-resolved output spectra were acquired to evaluate the spectral structure and bandwidth of masing under propagation.

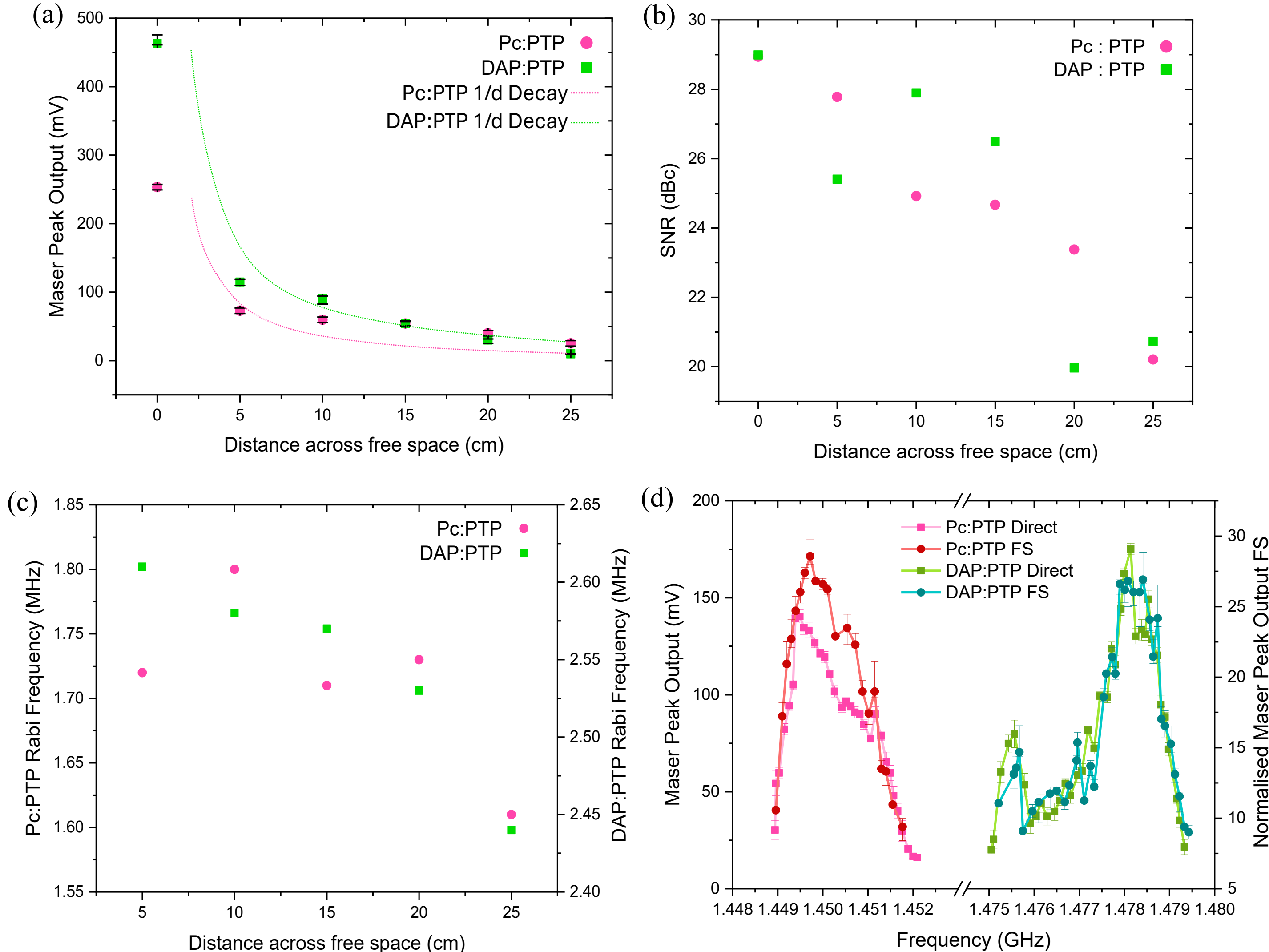

Figure 2: Baseline performance metrics. (a) Measured maser peak power of DAP: PTP organic sample as a function of the distance between the transmitting patch antenna and receiving probe, recorded in 5 cm increments from the direct connection (0 cm) to 25 cm in FS. At 0 cm, the signal is connected directly to the oscilloscope. The dashed curves show reference $1/d$ decay trends beginning at 2 cm, representing expecting attenuation from FS propagation without environmental losses. (b) SNR of the maser signal across the FS distances. SNR values were obtained directly from the Picoscope software, which calculates SNR by comparing the peak spectral amplitude to the surrounding noise floor. A gradual decline in SNR is observed with distance, yet a prominent signal persists across the entire range tested. The lowest recorded SNR, at 20 cm, remains measurable at 18.7 dBc, demonstrating the robustness and detectability of the maser sign al over FS. (c) Rabi frequency as a function of FS transmission distance for Pc:PTP and DAP:PTP. Frequencies were extracted from the time-domain maser oscillations using transient signal analysis. (d) Frequency-resolved maser peak output for Pc:PTP and DAP:PTP under direct coaxial detection and 25 cm FS transmission. The resonance frequency was tuned via the mechanical plunger to sweep across masing transitions, and the output amplitude at each frequency point was obtained from the FFT of the oscilloscope. While absolute FS signal amplitude decreases with distance, the spectral line shape remains preserved, confirming that FS transmission maintains the underlying gain media characteristics.

A notable drop in maser peak power is observed between 0 cm (direct connection) and 5 cm separation, corresponding to the transition from guided signal delivery to radiative FS propagation. This region spans the reactive near-field (typically $< \lambda/2\pi \approx 3.4$ cm) and Fresnel zone, where the antenna radiation pattern is not fully formed. For the patch antenna used (patch width D ≈ 75 mm), the onset of the far-field region occurs at distances beyond approximately $r_{far} = 2D^2/\lambda \approx 5.5$ cm for λ ≈ 20.4 cm (1.47 GHz). This agrees with the measurement, where the $0 - 5$ cm region shows near/Fresnel-zone losses, while beyond 5 cm the attenuation trend stabilises into the expected FS path loss.

The maser peak power, SNR, and Rabi frequency gradually attenuate with distance, consistent with expected propagation losses, yet the signal remains clearly detectable at all separations. Single-shot SNR values remain above 20 dBc, indicating strong received signal quality.

The observed Rabi oscillations in Figure 2c exhibit a slight lengthening in period with distance, consistent with a reduction in local field amplitude as the transmitted power decreases. This behaviour arises from the power dependence of the driven Rabi frequency ($\Omega \propto \sqrt{n}$) rather than any change in intrinsic coupling strength. The frequency-domain response of this coupling is addressed in Section 2.5, where the corresponding normal-mode splitting is examined.

At 25 cm, both materials sustain their full spectral tuning ranges: 1.449–1.452 GHz for Pc:PTP and 1.475–1.479 GHz for DAP:PTP, corresponding to total masing bandwidths of approximately 3 MHz and 4.3 MHz, respectively. The central masing frequencies ($f_0 = 1.449$ GHz for Pc:PTP and $f_0 = 1.478$ GHz for DAP:PTP) represent the points of maximal gain and frequency stability and remain accessible even at the maximum tested FS separation. This confirms that both systems retain sufficient output strength to support complete spectral tuning and coherent operation under propagation.

## 2.3 Directionality and Angular Alignment Sensitivity

Precision alignment presents persistent challenges for free-space optical (FSO) and quantum key distribution (QKD) links where even small misalignments elevate error rates and compromise security [19] [20]. This challenge is particularly critical in satellite communication, autonomous sensing, and quantum data transfer, which demand beam fidelity and low error rates. Microwave links, by virtue of their longer wavelengths and broader beam divergence, offer greater tolerance to angular offsets. However, conventional sources lack the coherence needed for high-fidelity applications, encouraging the evaluation of maser signals under controlled angular perturbations.

To evaluate the maser-based system's directional characteristics and alignment sensitivity, the receiving patch antenna was mounted on a 360° azimuthal rotation stage and incrementally rotated while maintaining a fixed distance from the transmitter. Measurements were recorded at 5 cm intervals between 5 cm and 25 cm, enabling characterisation of angular dependence in signal strength and spectral stability. This experiment provides a controlled point sensitivity test by varying the receiver's azimuth angle at fixed separation, enabling quantitative characterisation of angular dependence in signal strength and spectral stability. Although not a complete model of satellite motion or multi-axis pointing errors, where pointing deviations of $< 1\,mrad$ can cause significant loss or link failure [21], such controlled misalignment tests are directly relevant to evaluating robustness against alignment perturbations in FS links. Microwave patches, with beam widths on the order of tens of degrees (77.5 deg in this paper) and intrinsic phase coherence, are far more tolerant of misalignment while preserving quantum-coherent maser features.

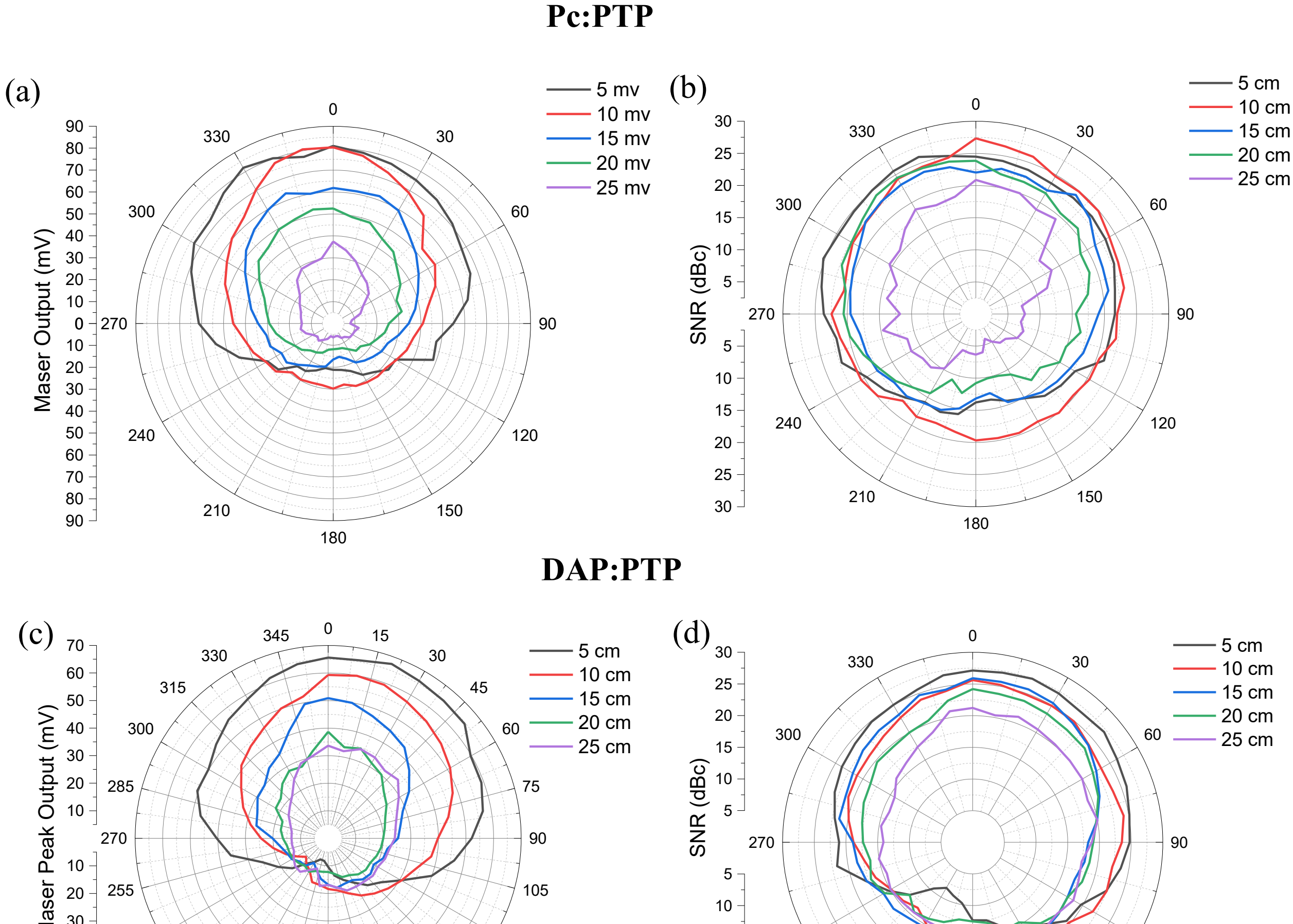

Figure 3: Angular radiation characteristics of masing emission under FS propagation. (a) Polar plot of Pc:PTP maser peak output voltage (mV) as a function of receiver angle, measured at distances of 5-25 cm. (b) Corresponding signal SNR of Pc:PTP at the same distances and receiver angles. (c) DAP:PTP maser peak output voltage (mV) measured across the same angular range and propagation distances. (d) SNR of DAP:PTP maser signal showing angular emission profiles at each distance.

The observed angular dependence of the maser signal confirms the directional radiation characteristics of the custom-designed patch antenna, with a measured 3 dB bandwidth of 77.5°, consistent with classical microstrip patch antenna theory [22]. Polar measurements of the received maser signal at various FS distances (5-25 cm) show a clear peak in signal strength when the receiver patch is aligned face-on with the transmitter and a consistent decrease in amplitude as the receiver is rotated away from this alignment. This behaviour reflects the expected forward-lobe radiation of the microstrip antennas. Laboratory reflections and multipath cannot be excluded; nevertheless, the dominant trend matches the predicted broadside emission profile, with reduced coupling at off-axis angles.

While signal levels were measurable across the full 360° rotation, this potentially arises from minor back-lobe radiation, edge diffraction, and residual reflections in the laboratory environment. These effects mean the polar plots should also be interpreted as a characterisation of the antenna system under experimental conditions, rather than a complete measure of maser-field angular robustness. Nevertheless, the maser signal remained detectable across all orientations and distances, with SNR reported, showing that antenna-limited behaviour still supports coherence in subsequent FS tests.

## 2.4 Environmental Resilience of Maser Transmission

Atmospheric scintillation and turbulence present significant challenges for FS communication systems, often causing signal fading, phase distortion, and coherence degradation [23] [24]. While FS microwave propagation in the L-band is generally resilient to such effects compared to optical systems, the practical stability of quantum-coherent maser signals must be empirically verified under realistic environmental conditions. Unlike conventional RF systems, masers encode information not only in amplitude but also in quantum-coherent properties such as the Rabi oscillation period and spectral linewidth, which are intrinsically determined by spin-photon coupling strength and collective interactions as described by the Tavis-Cummings model [13].

This section evaluates whether the measured maser power remains consistent with classical FS attenuation models, confirming that any observed losses originate from propagation effects rather than the intrinsic instability of the maser source.

To evaluate the environmental robustness of the maser signal, a series of controlled FS experiments under two representative real-world stressors was conducted: high atmospheric humidity and partial foliage using hydrated plant leaves. Peak maser output (mV) was measured under each condition and compared against Friis' transmission predictions, providing a benchmark for distinguishing natural attenuation from coherence-related degradation. The Friis transmission equation is denoted as:

$$\frac{P_r}{P_t} = G_t G_r \left(\frac{\lambda}{4\pi d}\right)^2 \qquad (1)$$

where $P_t$ and $P_r$ are the transmitted and received powers, $G_t$ and $G_r$ are the unitless gains of the transmitting and receiving antennas, $\lambda$ is the maser wavelength, and $d$ is the separation between antennas. In our measurements, the detector output was recorded in millivolts, which is proportional to the received electric field amplitude rather than power. Since electric field amplitude scales as the square root of received power, the Friis $1/d^2$ dependence reduces to an inverse-distance $(1/d)$ trend for our voltage data. The dashed $1/d$ traces in Figures 4a and 4b were scaled to the measured signal at the shortest distance (5 cm) and served as a benchmark for expected FDS attenuation in the absence of additional losses. Departures from this trend indicate potential attenuation due to humidity or foliage-related absorption.

In the first experiment, a sealed humidity enclosure was built around the 5 - 25 cm FS transmission path. Maser signals were measured at two relative humidity extremes $< 40\%$ and $\sim 90\%$ using a digital hygrometer (RS PRO 2216358). The source cavity and patch transmitter/receiver remained fixed throughout, and coaxial feedlines were disconnected to ensure the FS link passed entirely through the controlled environment.

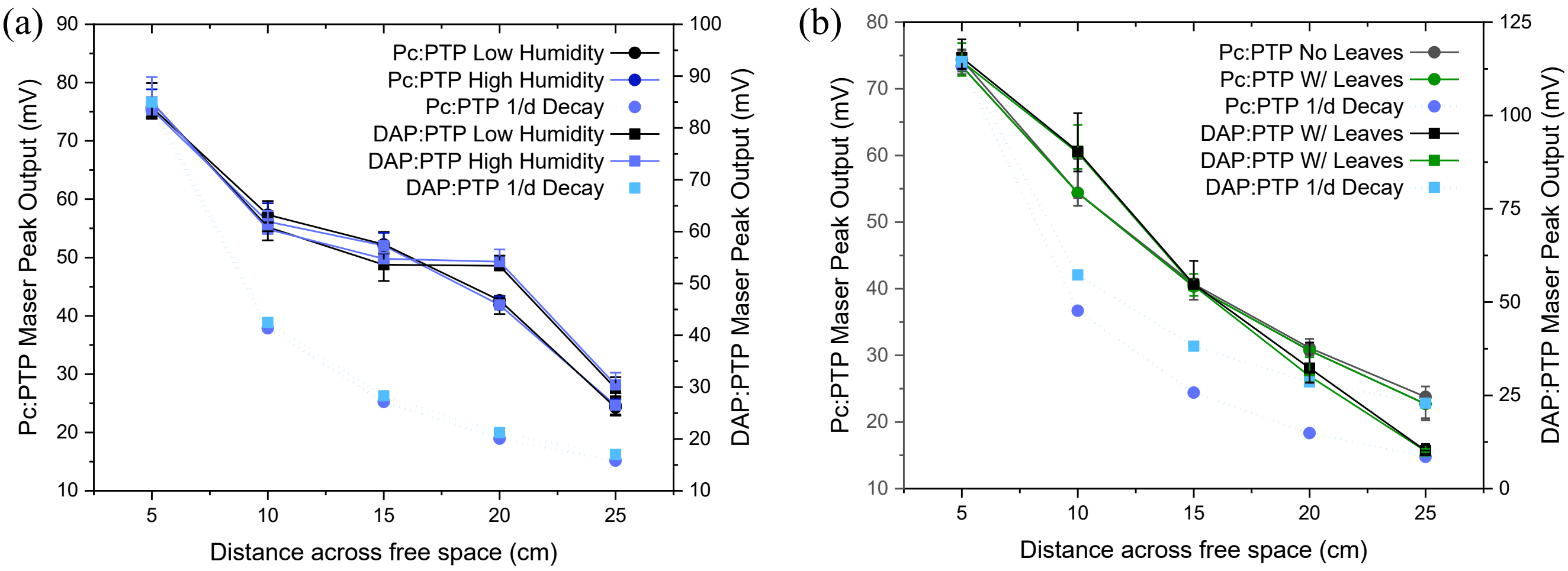

Figure 4: Maser signal analysis with environmental interference. The dashed $1/d$ curves are reference benchmarks derived from the Friis transmission relation, scaled to the measured signal at 5 cm. As the detector output is proportional to the received electric field amplitude (rather than power), the Friis $1/d^2$ dependence for power reduces to a $1/d$ trend for the measured voltage. These lines therefore represent the expected FS attenuation without additional environmental losses. (a) Maser peak output as a function of transmission distance, comparing low (<40%) and high (>90%) humidity conditions for both Pc: PTP and DAP: PTP gain media. (b) Maser peak output as a function of transmission distance with and without vegetation, tested under low-humidity conditions (<40%).

For the second experiment, the receiver antenna was fully occluded at each distance point using freshly cut laurel leaves. Care was taken to maintain consistent leaf hydration and surface coverage to isolate the dielectric/scattering effects from ambient changes. These tests provide a rigorous baseline for assessing how coherent maser features persist through materials that typically scatter or depolarise conventional signals.

Figure four shows Pc:PTP exhibited deviations of <1% in maser output under humidity and ~1.3% with vegetation, while DAP:PTP showed shifts of ~0.8% and ~2%, respectively. These variations are comparable to typical shot-to-shot fluctuations in optical pumping and alignment and therefore should not be taken as direct evidence of foliage transparency. The Rabi frequencies deviated by ~3% for Pc:PTP at isolated distances, though average changes across both materials and conditions remained below 1%, with no signs of cumulative degradation over distance or time. It is also noted that minor fluctuations in optical pump power may have contributed to the observed percent-level variations.

## 2.5 Quantifying Strong Coupling and Rabi Splitting in Emission

Cavity quantum electrodynamics (cQED) enables coherent energy exchange between matter and quantised electromagnetic fields, with the strength of this interaction governed by the spin-photon coupling rate, $g$ [7]. In the strong coupling regime, coherent oscillations occur between the spin ensemble and a cavity mode faster than their decay rates, producing coherent oscillations between the two systems at the normal mode splitting frequency $\Omega$. Achieving and quantifying this regime in the present maser platform was made possible by the high peak output power of the device, which enables a large intracavity photon population $\langle a \dagger a \rangle$ and, consequently, observable Rabi dynamics under free-running emission. The experimental signatures of this regime appear as Rabi oscillations in the time domain and corresponding normal-mode splitting in the frequency domain, which establish the foundation for coherent microwave maser emission.

To experimentally validate the presence of strong coupling and coherent energy exchange, we directly recorded the maser emission in the time domain using a 50 Ω terminated digital oscilloscope. The resulting waveform (Fig. 5a) reveals clear and well-resolved Rabi oscillations, reaching a peak maser output of 463 mV, corresponding to approximately 4.29 mW or +6.32 dBm, surpassing the previous highest reported pulsed maser peak power to date (2.31 mW or +3.69 dBm) [18].

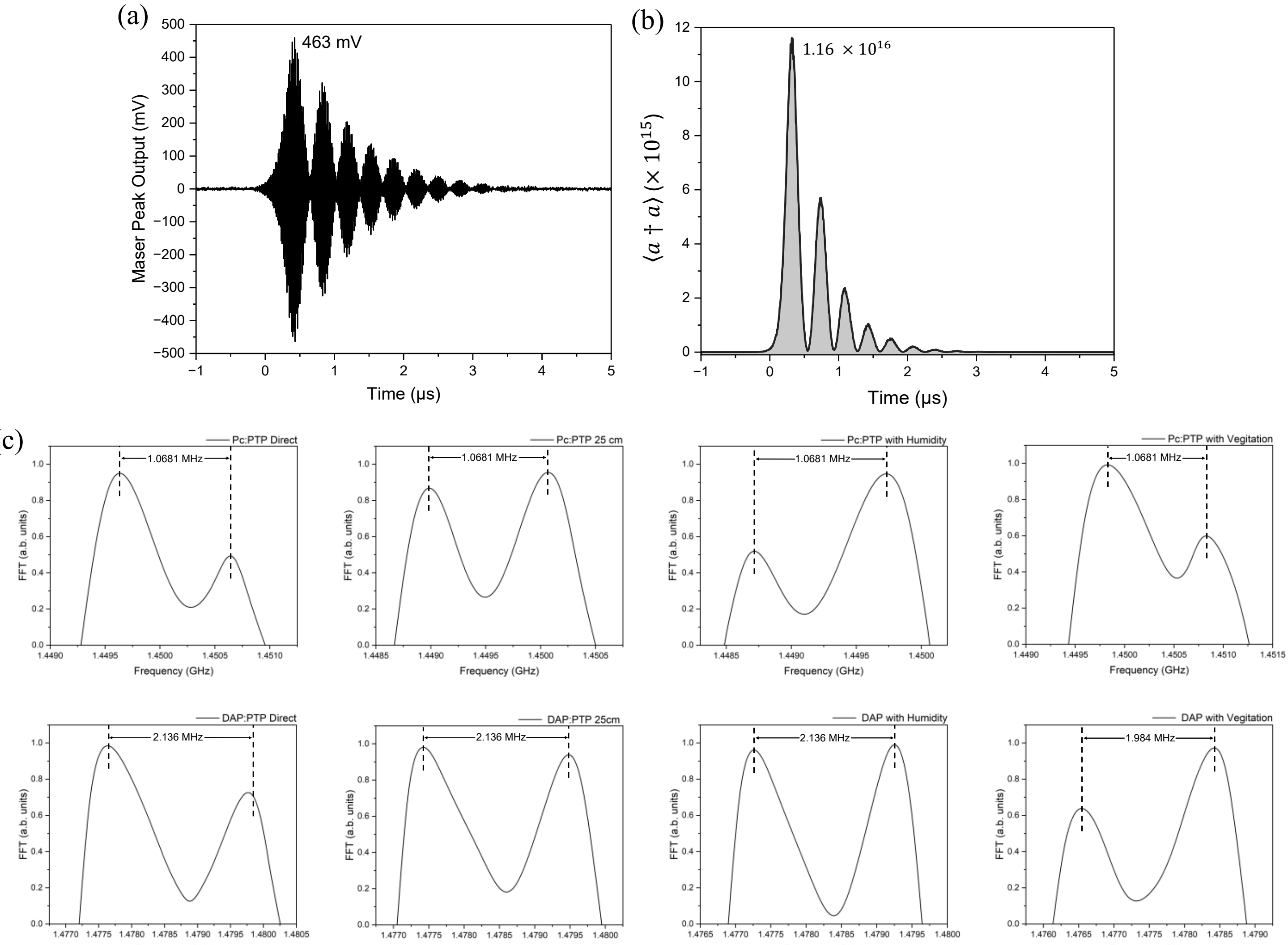

Figure 5: Quantifying strong coupling in the frequency and time domain. (a) Time-domain maser emission captured directly by a digital oscilloscope with a $50\,\Omega$ input termination, following direct coaxial connection to the maser output. The trace shows Rabi oscillations with a peak voltage of 463 mV, corresponding to an instantaneous output of 4.29 mW (+6.32 dBm). (b) Intracavity photon number $\langle a \dagger a \rangle$ for the DAP:PTP direct-detection case, obtained from the maser equation using $V_{rms}$. The peak value is approximately $1.16 \times 10^{16}$ photons. (c) Frequency-domain vacuum Rabi splitting observed via FFT analysis of maser transients for Pc:PTP (top) and DAP:PTP (bottom) under four conditions: direct coaxial detection, 25 cm transmission, elevated humidity, and vegetative occlusion. The measured splitting of 1.0681 MHz (Pc:PTP) and 2.136 MHz (DAP:PTP) remain spectrally invariant to a high number of significant figures across all environments. FFT traces were smoothed using a Bézier interpolation; this did not alter the extracted frequency separation values.

To quantify the extent of spin-photon interaction in the FS maser system, we evaluate the cooperativity $C^*$, defined as:

$$C^* \approx \frac{4g_e^2}{\kappa_c \kappa_s} \qquad (2)$$

where $g_e$ is the single-spin coupling constant, $\kappa_c$ is the cavity decay rate, and $\kappa_s$ is the spin decoherence rate. Strong coupling is achieved when $C^* > 1$, where stimulated emission dominates over dissipative losses.

In this study, we adopt $g_e = 2\pi \times 2.3$ MHz, consistent with values reported previously for 0.01% DAP:PTP masers [30]. The spin decoherence rate was taken as $\kappa_s = 2/T_2^*$, where $T_2^*$ is the inhomogeneous dephasing time. Reported values for $T_2^*$ for organic pentacene triplets span $1.1 - 2.9$ µs [30], [31]. The cavity decay rate is determined by $\kappa_c = \omega_c/Q$, with $\omega_c = 2\pi f_c$. With $Q = 8000$ for both Pc:PTP and DAP:PTP devices, and resonant frequencies of $f_c^{Pc} \approx$ 1.45 GHz and $f_c^{DAP} \approx 1.478$ GHz, we find $\kappa_c^{Pc} = 2\pi \times 0.181$ MHz and $\kappa_c^{DAP} = 2\pi \times 0.185$ MHz. Substituting these values yields cooperativity ranges of $C^* \in [403 - 1063]$ for Pc:PTP and $C^* \in [395 - 1043]$ for DAP:PTP.

These results confirm that both systems operate deeply within the strong-coupling regime. Notably, the extracted values exceed those reported in the literature for room-temperature organic masers, where Breeze *et al.* reported $C^* \approx 190 - 250$ for Pc:PTP, and Ng *et al.* reported $C^* \approx 182$ for DAP:PTP. The results therefore represent an enhancement by a factor of $\sim 1.6 - 4.2$ (Pc:PTP) and $\sim 2.2 - 5.9$ (DAP:PTP) relative to prior state-of-the-art values, underscoring the cavity-mediated spin–photon coupling efficiency in this platform.

To further quantify the strength of stimulated emission, we evaluate the intracavity photon population, $\langle a \dagger a \rangle$, from the directly detected maser transients. The voltage trace is converted to output power using $P(t) = V_{RMS}^2/50$, so evolves consistently with the measured time-domain output with:

$$\langle a \dagger a \rangle = \frac{P(t)(1+K)}{hf_{mode}\kappa_c K} \qquad (3)$$

where $P(t)$ is the peak maser output power, $f_{mode}$ is the cavity resonance frequency, and $K$ is the coupling coefficient (taken as 0.20). For the DAP:PTP sample, this yields a peak intracavity photon number of $1.16 \times 10^{16}$ photons. The temporal evolution of $\langle a \dagger a \rangle$ follows the observed maser envelope, with successive oscillations mirroring the coherent Rabi dynamics visible in the time-domain voltage trace. These values exceed earlier reports of organic maser intracavity populations [18], reflecting the heightened maser power. Further evidence of strong spin-photon coupling is provided by normal-mode splitting estimation, resolved in the frequency domain via FFT analysis (Fig. 5c). In the Travis-Cummings framework, the angular normal mode splitting $\Omega = 2g_e = 2g\sqrt{N}$ represents the rate of exchange between the spin-ensemble and cavity field.

Pc:PTP and DAP:PTP exhibit persistent, symmetric splitting under direct coaxial detection through to 25 cm of FS propagation including two environmental perturbations: high humidity and partial foliage occlusion. The measured splitting values, 1.0681 MHz for Pc:PTP and 2.136 MHz for DAP:PTP, remain consistent from direct detection and across FS conditions, including elevated humidity and vegetative occlusion. A slight carrier-frequency shift observed in the DAP:PTP foliage case is consistent to modest thermal pulling of the source; notably, the Rabi splitting remains unchanged. Notably, these values reproduce those reported in our previous cavity-coupled characterisation [18], where FFT analysis yielded 1.37 MHz and 2.14 MHz for Pc:PTP and DAP:PTP, respectively. This distinctive agreement underscores the spectral stability of vacuum Rabi splitting under FS propagation and reinforces the conclusion that strong coupling is maintained at the source.

The Rabi frequencies obtained from the time-domain maser bursts ($\approx$ 1.7 MHz for Pc:PTP and $\approx$ 2.6 MHz for DAP:PTP) are higher than the splittings observed in the frequency domain (1.07 MHz and 2.14 MHz, respectively). This difference can be attributed to the large number of excitations in the cavity during masing. When the intracavity photon population becomes large, the Rabi frequency scales approximately as $\Omega_{Rabi} \propto \sqrt{n_{cavity}}$ , leading to an enhanced oscillation rate. This observation signifies a transition from the quantum to classical regime, where the microwave field behaves as a classical drive for the spin ensemble.

## Discussion

This study presents a proof-of-concept demonstration of a temporal coherence maser emission maintained during FS propagation of a room-temperature maser signal, The transmission maintains stable Rabi oscillations and consistent normal-mode splitting over distances up to 25 cm, approximately one wavelength at the masing frequencies of 1.45–1.48 GHz. The results verify that the high-Q dielectric cavity can sustain coherent spin–photon interactions internally while generating a radiative maser field that retains those coherence properties during propagation. This stability is enabled by the milliwatt-level DAP:PTP and Pc:PTP maser sources, which provide sufficient field strength to overcome near-field loss and validate coherence preservation under realistic transmission conditions.

Peak output measurements reveal a considerable drop between direct coaxial connection and 5 cm propagation, as expected during the transition from guided to unguided radiation. Beyond this, the received power levels are in close agreement with Friis' transmission law, which provides a theoretical baseline for attenuation in ideal FS conditions. This agreement confirms residual losses arise primarily from propagation effects rather than intrinsic decoherence.

The transmission and environmental tests were conducted on different days and following varying durations of optical pumping; therefore, the key comparison is between the conditions applied to each crystal type individually. For both Pc:PTP and DAP:PTP maser signals, the introduction high humidity (>90%) or foliage occlusion resulted in only minimum signal degradation. These minor differences are consistent with expected experimental variability, including slight fluctuations in optical pumping output. The measurements represent preliminary but representative evaluations of environmental robustness, where each data point is averaged over five or more maser signals to reduce the impact of natural optical fluctuations.

As anticipated based on their distinct crystal properties, Pc:PTP and DAP:PTP exhibit differences in absolute peak output, and SNR values. However, in all cases the coherence of transmission is maintained; time-domain Rabi oscillations (Fig. 2c) and frequency-domain normal-mode splitting (Fig.5c) persist under distance, humidity and foliage stressors, providing direct evidence that spin–photon coherence is preserved during radiative propagation. This confirms that the maser's observed Rabi frequency/normal mode splitting is determined by the spin–photon coupling strength rather than environmental factors, and is sustained in FS, validating that the process responsible for temporally coherent masing is retained outside the resonator.

The normal-mode splitting remains constant to high precision: 1.068 MHz (Pc:PTP) and 2.136 MHz (DAP:PTP), are comparable with resonator-bound values reported previously [18].

Complementary time-domain measurements show coherent Rabi oscillations at $\approx 1.7$ MHz (Pc:PTP) and $\approx 2.6$ MHz (DAP:PTP), verifying that the coupling dynamics persist during masing and demonstrating that the spin–photon interaction remains robust under FS operation. These magnitudes are also consistent with literature reports of normal-mode splittings around 1.8 MHz for Pc:PTP masers [25], confirming that the observed spin-photon hybridisation lies within the established strong-coupling regime. This direct correspondence establishes that the spin–photon coupling strength remains unchanged even when the maser output is radiated into FS.

These features are often lost in conventional microwave or optical links when exposed to turbulence, foliage-induced attenuation and alignment errors. For example, empirical studies of radio-wave propagation through tree canopies show that the signal's coherent (direct) component decays rapidly with foliage depth, causing the scattered/incoherent component to dominate and degrade phase and amplitude stability [26]. Whereas, in this work, the maser signal remains measurable under elevated humidity (>90% RH), polarisation misalignment, and even biological attenuation using hydrated vegetation. This resilience contrasts conventional FSO and QKD systems, where misalignment and scattering severely degrade coherence and link fidelity [2] [19] [20].

Further, among the highest reported at room temperature, a record-high maser peak output of 463 mV was observed under direct coupling, corresponding to 4.29 mW (+6.32 dBm) into a 50 Ω load. This result exceeds previously reported room-temperature maser powers [27] [18] and sets a new benchmark for solid-state coherent microwave emission. The corresponding intracavity photon population, estimated at $1.16 \times 10^{16}$, also ranks among the highest directly detected for a room-temperature solid-state maser. The Pc:PTP device also operates in the milliwatt regime, meaning that dual-frequency operation of Pc:PTP and DAP:PTP could provide side-by-side coherent maser transmission at distinct frequencies, a promising forward-looking route if further power scaling is achieved.

This behaviour parallels the operational principle of the hydrogen maser atomic clock, the frequency standard, where a high-Q microwave cavity sustains atomic coherence and a lower-power signal is extracted for precision timing. Similarly, the present system retains the stabilising influence of the dielectric resonator while exploiting its high peak output to propagate a coherent field into FS. Unlike the hydrogen maser, which operates in the sub-μW regime, the milliwatt-level Pc:PTP maser delivers sufficient field strength for direct transmission without external amplification.

Even at the maximum 25 cm separation, SNR levels remain above 20 dBc with direct line of sight, and this performance is retained across a 90° angular misalignment. The 5 MHz bandwidth corresponds to the spectral width measured from the FFT of the time-domain maser signal, while the −100 dBm (0.1 pW) noise floor reflects the consistent baseline level observed in the same FFT spectra. Using the Shannon-Hartley theorem, the channel capacity is given by:

$$C = B \log_2(1 + SNR) \approx 140 \text{ Mbps} \qquad (4)$$

where $B = 5$ MHz, and the FS output is 40 mV into 50 Ω. This estimate underscores the maser's potential for high-rate, narrowband communication with quantum-level phase stability, an enticing alternative to bulky cryogenic systems or environmentally fragile optical links.

While FS amplitudes are reduced at distance, the spectral structure is retained, preserving channel identity essential for quantum or classical information encoding. Following this success, the current platform could be improved further. As the 532 nm pump laser overlaps with the gain media's smallest absorption spectra, and the patch antennas that were milled in-house deliver modest reflection performance ($S_{11} \approx -16\,dB$) and directionality. These constraints provide a conservative baseline. Planned upgrades include the use of LED-pumped luminescent concentrators (LED-LC) and precision-fabricated transceivers would drive higher maser output, expanding link range and data capacity.

## Conclusion

This work demonstrates that cavity-based masers can sustain temporally coherent emission beyond resonant boundaries, achieving FS transmission without waveguides. Retaining key quantum features such as Rabi oscillations and consistent normal-mode splitting throughout radiative propagation confirms that spin–photon coherence generated within a high-Q dielectric cavity persists into the emitted maser field. This represents a essential step in validating masers as viable coherent microwave sources capable of operation outside enclosed resonator systems.

Environmental validation further revealed that this coherence is resilient to atmospheric humidity (>90%), dielectric occlusion, and moderate angular misalignment, indicating that observed losses arise primarily from propagation rather than intrinsic decoherence. The maser signal's amplitude followed the expected FS attenuation profile while maintaining spectral integrity, confirming that temporal coherence can survive realistic environmental conditions over short distances.

The observed milliwatt-level peak output enabled FS maser operation with a high signal-to-noise ratio (>20 dBc at 25 cm) at room temperature. This marks a significant advance toward scalable, high-spectrally pure microwave emitters. In many respects, this progression mirrors the evolution of the hydrogen maser from laboratory experiments to a practical frequency standard. However, this study demonstrates temporal coherence retention under ambient conditions using a solid-state gain medium with high peak power.

In the near term, compact room-temperature masers with directional FS links offer practical solutions for quantum sensing technologies where coherence, not link reach or power, is paramount. The demonstrated platform could support quantum radar calibration, precision timing distribution, and low-power coherent control in scalable quantum devices and sensor networks. This study thus reframes masers from niche scientific instruments into emergent, practical transmitters, with potential applications in metrology and secure communications.

# Methods

*Experimental Setup*

The maser cavity was designed using a commercially available finite-difference time-domain (FDTD) simulator (CST Studio Suite®). The cavity comprises a cylindrical copper with an inner diameter of 40 mm and a height of 35 mm, loaded with a strontium titanate (STO) dielectric annulus resonator. The STO resonator dimensions are 10.12 mm (outer diameter), 3.05 mm (inner diameter), and 14.52 mm (height), and supports a $TE_{01\delta}$ mode aligned with the crystal y-axis to optimise overlap with the spin ensemble.

A single organic crystal (~3 mm in diameter and 5-7 mm in length), either pentacene-doped p-terphenyl (Pc:PTP) or diazapentacene-doped PTP (DAP:PTP), was placed at the centre of the annulus. The crystal-resonator assembly was supported on a PTFE pedestal 2 mm above the cavity base for field uniformity above the copper base.

Stimulated microwave emission was extracted via a coaxial loop antenna at the cavity end port. The resonance frequency $f_0$ was tuned via a mechanical plunger to match the masing transitions of the respective gain media: 1.4496 GHz for Pc:PTP and 1.4783 GHz for DAP:PTP. Emission was measured directly at the receiver antenna using a 50 Ω-matched digital oscilloscope without amplification or filtering to preserve waveform fidelity. Instantaneous output power was calculated from the peak maser voltage $V$ using $P = \frac{V^2}{R}$ for example, 463 mV corresponds to 4.3 mW (+6.3 dBm)

Time-domain traces were recorded using a PicoScope 6000 series oscilloscope at a sampling rate of 1 GS/s. These traces captured the natural oscillatory decay of the maser output following optical excitation. Rabi oscillation periods were extracted by measuring the time between successive peaks in the maser envelope. The corresponding Rabi frequency was computed via equation 2.

A cylindrical copper cavity houses the $SrTiO_3$ (STO) dielectric resonator, supporting the $TE_{01\delta}$ mode, with the resonance frequency finely tuned via a mechanically adjustable plunger. The resonator contains the organic gain media (Pc: PTP or DAP: PTP), which is optically excited using a pulsed 532 nm Nd: YAG laser introduced laterally. Optical pumping induces population inversion in the gain medium, enabling coherent maser emission. The maser signal is extracted via a coaxial loop antenna in the copper cavity and projected into FS using a linearly polarised patch antenna designed on CST Studio Suite and milled in-house. A second, linearly polarised patch antenna, a copy of the transmitter, receives the transmitted signal, which is then recorded using a digital oscilloscope (Picoscope 6428E). The central 3D far-field plot (simulated in CST Microwave Studio) illustrates the directional radiation pattern of the patch antenna, with the polarisation axis and main lobe highlighted. The far-field plot uses a green-to-red colour scale, where red corresponds to regions of the highest directionality and signal intensity, and green indicates lower intensity.

*Purcel Factor*

A key parameter in accessing this regime is the Purcell factor, which scales with the resonator's quality factor ($Q$) and inversely with its effective mode volume ($V_m$). In our system, a high-permittivity $SrTiO_3$ dielectric resonator ($\varepsilon_r \approx 318$) yields low $V_m$, resulting in magnetic Purcell factors exceeding $10^8$ [28]. This enables strong coupling between the triplet spin ensemble in the organic gain medium and the confined microwave field, even at room temperature. When embedded in a high-Q ($Q \approx 8000$) microwave cavity, both Pc:PTP and DAP:PTP demonstrate strong masing with strong spin-photon interaction.

*Free-Space Transmission and Environmental Testing*

The maser signal was broadcast into FS using a pair of linearly polarised, single-layer rectangular microstrip patch antennas, designed and fabricated in-house to operate within the masing frequencies of 1.45-1.48 GHz. Each patch consisted of a copper top layer of dimensions approximately 74.91 mm (width) × 47.55 mm (length), mounted on an FR4 substrate ($\varepsilon_r = 4.3$, 1.6 mm thickness), with a rear copper ground plane. The antennas were fed via an impedance-matched microstrip feedline (50 Ω), with dimensions of 3 mm width and 29.1 mm length, optimised to for minimal return loss and well-matched transmission across the L-band masing window.

Both the transmitter and receiver antennas were mounted on independently adjustable linear translation stages, allowing smooth and precise control of separation in the x-direction (to millimetre resolution). To assess angular misalignment effects, the receiver antenna was mounted on a separate rotation stage and swept azimuthally (θ = 0-360°) about the transmission axis.

For environmental testing, a transparent acrylic chamber (open at the base) was placed over the optical table, enclosing the transmission path. A commercial ultrasonic nebuliser introduced water vapour via a directed hose positioned above the antenna aperture. Relative humidity levels exceeding 90% were achieved and monitored throughout using a digital hygrometer (RS PRO RS-91, ±3% RH accuracy, 0-100% range).

To simulate atmospheric occlusion and dielectric scattering, freshly cut laurel-type leaves was suspended vertically between the antennas, fully intersecting the transmission axis, directly covering all the receiver antenna. In all test conditions, no adjustments were made to the antenna alignment or pumping configuration, isolating the effects of misalignment, moisture, and occlusion on maser signal integrity.

*Optical Pump*

Optical excitation of the masing medium was provided by a pulsed 532 nm laser, delivering ~5 ns pulses at a fixed repetition rate of 1 Hz. The incident pulse energy was held constant at 30 mJ for all experimental conditions. A collimated beam with a 2.5 mm spot diameter was directed perpendicularly onto the top surface of the organic crystal, ensuring uniform excitation across the active region. Each crystal was centrally positioned within the bore of the strontium titanate (STO) annulus resonator to maximise spatial overlap between the optical pump region and the resonant microwave field.

*Vacuum Rabi Splitting and Strong Coupling Assessment*

The frequency-domain vacuum Rabi splitting **Ω**, observed as two spectrally resolved peaks, is a key hallmark of strong coupling in cQED. The splitting magnitude is defined by:

$$\Omega = 2g_e = 2g\sqrt{N} \qquad (5)$$

where is the single-spin coupling constant and is the number of active spins in the masing ensemble. In this study, vacuum Rabi splitting of 1.0681 MHz (Pc:PTP) and 2.136 MHz (DAP:PTP) were extracted analysis across direct, FS, and environmentally perturbed conditions.

*SNR*

SNR values were obtained directly from the PicoScope 6428E FFT analysis. According to the PicoScope User Manual [29] SNR is defined as:

$$SNR = 20log_{10}\left(\frac{RMS\ value\ of\ carrier\ ('datum')}{\sqrt{\sum of\ squares\ of\ all\ values\ excluding\ datum\ and\ harmonics}}\right)$$

where the 'datum' corresponds in this work to the carrier peak at the masing transition frequency.

Rabi Frequency

To probe this regime experimentally, we extract the Rabi frequency ($f_{Rabi}$) by measuring the oscillation period in the time-domain maser signal, using the relation:

$$f_{Rabi} = \frac{2\pi}{T_{Rabi}} \qquad (6)$$

where $T_{Rabi}$ is the time between successive maser oscillation peaks.

*Crystal growth*

Crystals of Pc:PTP and DAP:PTP used in this study were synthesised via the Bridgmann technique, as described previously [30] . Doped powders were prepared by thoroughly grinding 0.01% molar ratios of either pentacene or diazapentacene with zone-refined p-terphenyl using a pestle and mortar. The mixtures were loaded into 4 mm inner-diameter borosilicate glass tubes (Hilgenberg GmbH) and sealed under an argon atmosphere. Crystal growth was achieved by passing the tubes through a furnace heated to 216 °C at a controlled rate of 4 mm/hr, yielding large cylindrical single crystals.